# Identify Web-page Content meaning using Knowledge based System for Dual Meaning Words

## Sukanta Sinha[1, 4], Rana Dattagupta[2], Debajyoti Mukhopadhyay[3, 4]

[1](TATA Consultancy Services, Victoria Park, Kolkata 700091, India)
[2](Computer Sc. Dept., Jadavpur University, Kolkata 700032, India)
[3](Dept. of Information Technology, Maharashtra Institute of Technology, Pune 411038, India)
[4](WIDiCoReL, Green Tower C- 9/1, Golf Green, Kolkata 700095, India)

**Abstract**

    **Meaning of Web-page content plays a big role while produced a search result from a search engine. Most of the cases Web-page meaning stored in title or meta-tag area but those meanings do not always match with Web-page content. To overcome this situation we need to go through the Web-page content to identify the Web-page meaning. In such cases, where Web-page content holds dual meaning words that time it is really difficult to identify the meaning of the Web-page. In this paper, we are introducing a new design and development mechanism of identifying the Web-page content meaning which holds dual meaning words in their Web-page content.**

## 1.  Introduction

    Web search engine is a tool that produces search results based on the user given query. World Wide Web (WWW) is a huge reservoir of Web-pages. Search engine crawler crawls down the Web-pages from WWW and creates a database of Web resources for the search engine [1, 2].

In the present era of Internet, WWW is an accumulated and interactive medium for accessing an enormous conglomeration of information [3]. The information in the Web-page content consists of diverse data types such as structured data, semi structured data and lack of structure of Web data, etc. [4]. Few cases we also found holds dual meaning words are exists in Web-page content. Meaning identification of those Web-page contents which holds dual meaning words is a challenging task.

The dual meaning word means a word which contains two meanings like 'bank' represents 'financial institute' as well as 'river side'. We need to identify the meaning based on the full sentence.

In our approach, we have mainly focused on identifying the Web-page content meaning, which holds only dual meaning words in their Web-page content. To identify the meaning, we have created a knowledge based system by collecting various types of data patterns.

Our paper is not intended to provide a complete survey of techniques. According to our knowledge, we have applied these techniques on few examples. Now a day's research on search engine has been carried out in universities and open laboratories, many dot-com companies. Unfortunately, many of these techniques are used by dot-coms, and especially the resulting performance, are kept private behind company walls, or are disclosed in patents that can be comprehended and appreciate by the lawyers. Therefore, we believe that the overview of problems and techniques that we presented here can be useful.

This paper discusses survey of the problem area in section 2. Section 3 discusses about the XML schema. Section 4 depicts the proposed approach. Section 5 shows some experimental analyses. Finally, section 6 concludes the paper.

## 2.  The Problem Area

    Web-page content meaning identification is an essential part of a search engine to produce relevant search result. Most of the cases we can get the Web-page content meaning from title or meta-tag area of that Web-page content but they do not always match with the actual Web-page content. On the other hand, a few cases where Web-page content holding dual meaning words are really difficult to identify the meaning of the Web-page content.

In general, our main goal is to identify the Web-page content meaning which holds dual meaning words in their Web-page content. The example illustrates the difficulty to identify the meaning of a Web-page content, which can be overcome by using our proposed system.

Example 1: John is looking for a *bank* to open a savings account on the other hand Alex is looking for a *bank* of the river for a get together. Here, both the *bank* represents different meaning, one for financial institutes and other one for river side. If both the sentence exists in different Web-page content then the meaning of the Web-page content need to be retrieved based on their content.

Example 2: Peter found a *bank* which located on the *bank* of the river. This is a single sentence which represents financial institutions as well as river side. This time any one of the meanings is valid for the sentence. In our approach, we assumed that one Web-page has only one meaning. Hence, for this type of situation we will assign any one meaning based on our programming logic.

## 3.   XML Schema

An XML Schema describes the structure of an XML document [5, 6]. The XML Schema language refers to an XML Schema Definition (XSD). The purpose of an XML Schema is to define the legal building blocks of an XML document. An XML Schema defines elements, attributes that can appear in a document [7, 8]. It also expressed data types, default and fixed values for elements and attributes. One of the greatest strengths of XML Schemas is the support for data types and written in XML. XML Schemas are extensible because they are written in XML.

XML Schema holds simple and complex elements [9, 10, 11]. A simple element is an XML element that contains only text. It cannot contain any other elements or attributes. A complex element is an XML element that contains other elements and/or attributes. There are four kinds of complex elements; they are empty elements, elements that contain only other elements, elements that contain only text, elements that contain both other elements and text. The <schema> element is the root element of every XML Schema. The <schema> element may contain some attributes [12, 13, 14].

## 4.   Proposed Approach

In our approach, we have proposed a mechanism which identifies meaning of Web-page content for those who holds dual meaning word in their Web-page content. Section 4.1 explains an overview of creating knowledge based system and section 4.2 depicts our algorithm.

### 4.1. Knowledge Based System Generation

To create a knowledge based system we have collected dual meaning words from various sources like internet, dictionary, etc. Now for each dual meaning word, we have created one XML which link with Fig.1 given XSD. The considered XSD holds simple and complex type of elements.

'dualMeaningWordName' attribute holds the dual meaning word name. 'keywords' is a complex element which holds various sets of keyword, which classified based on their meaning. 'keyword' also a complex type element which holds similar types of key elements with their meaning. 'names' is a complex type element which holds key element names that represent same meaning. 'name' and 'meaning' are simple type element holds key values and their meaning. Each XML holds a 'dmw_id'. We have maintained dual meaning word with a corresponding 'dmw_id'. Key words are taken from dual meaning word holding meaning. For example "John is looking for a bank to open a savings account" and "Alex is looking for a bank of the river for a get together" holds 'account', 'river' key words. All the key word meaning is taken care while design the XML. In Fig.2 we have shown a part of an XML for 'bank'.

**figure 1.** A sample XSD

**figure 2.** A part of an XML (for bank)

### 4.2. Algorithm

To identify Web-page content meaning we are using below given algorithm. This algorithm mainly focused on identifying the Web-page content meaning, which holds dual meaning words in their Web-page content. In our approach, we have used a knowledge based system for identifying the meaning of dual meaning words. The knowledge based system stores the information in XML form.

```
Input : Web-page content
Output : Meaning of the Web-page
content

1. Extract dual meaning words from the
   Web-page content.
2. get count of dual meaning words in
   the Web-page content
3. if count = 0 then
     set isDualMeaningFlag:=False  and
exit
4. if count = 1 then
   a) set isDualMeaningFlag:=True
   b) Extract key words in the dual
      meaning word sentence
   c) Based on the key word traverse
      XML (knowledge based system) for
      dual meaning word
   d) Retrieve the meaning of that key
      and store it in a temporary
      table.
   e) Go to step 6.
5. if count > 1 then
   a) set isDualMeaningFlag:=True
   b) select the max occurred dual
      meaning word in the Web-page
      content
   c) if there exists multiple dual
      meaning word with same number of
      occurrence then select dual
      meaning word which appeared
      first in the Web-page content
   d) Extract key words in the dual
      meaning word sentence
   e) Based on the key word traverse
      XML (knowledge based system) for
      dual meaning word
   f) Retrieve the meaning of that key
      and store it in a temporary
      table.
   g) Go to step 6.
6. Choose the meaning from temporary
   table which count is maximized.
7. end
```

## 5.  Experimental Analysis

In this section, we have given some experimental study as well as discussed how to set up our system. Section 5.1 explains our experimental procedure, and section 5.2 shows the experimental results of our system.

### 5.1. Experimental Procedure

Performance of our system depends on various parameters and those parameters need to be set up before running our system. The considered parameters are Web-page repository, knowledge based system, i.e., dual meaning word XML with proper meaning, XML schema, etc.  Initially, we have created the knowledge based system with the help of internet, dictionary. Then we have tuned the knowledge based system through our experiments. In our experiment, we have taken a Web-page from our repository and pass it through our system and check the database for the meaning of that Web-page. If the Web-page holds dual meaning words then the meaning will identified otherwise update 'isDualMeaningFlag' as false.

### 5.2. Experimental Results

It is very difficult to compare our system with any existing system. Anyhow we have produced few data to measure our proposed system performance. As a part of experimental results, we have produced a statistic, which given in Table 1.

**Table1.** Performance Report of Our System

| No. of Web-page Taken / Repository Size | No. of Web-page hold Dual Meaning Words | Correct Meaning Identified in 1st Run | No. of XML Modified | Correct Meaning Identified after XML Modification |
|---|---|---|---|---|
| 1000 | 30 | 22 | 6 | 28 |
| 2000 | 50 | 43 | 5 | 47 |
| 3000 | 80 | 71 | 6 | 76 |
| 4000 | 110 | 99 | 9 | 104 |
| 5000 | 140 | 127 | 10 | 134 |

## 6.  Conclusion

Web-page content meaning identification is a very difficult job for any system. The human brain can find it easily but need to go through each and every Web-page contents, which is really impossible. We found that approximate 30% - 40% Web-pages are representing unique meaning; out of those 30% - 40% approximate 8% - 10% Web-pages are holding dual meaning words. Hence, we are concentrating to create those 8% - 10% Web-page meaning XML. We found approximate 95% successful cases achieved to identify Web-page content meaning those held dual meaning words in their Web-page content. Our approach is highly scalable. Suppose, we encountered a new pattern and want to support that pattern, then we just introduce the meaning XML and the system will work. We have tested our system by taking a sub-set of Web-pages shown in experimental results section. In this paper, we are mainly focused on our approach, which will work for large volume of data.